\documentclass[conference]{IEEEtran}
\IEEEoverridecommandlockouts

\usepackage[style=ieee, sorting=none]{biblatex}
\bibliography{refs} 

\usepackage{amsmath,amssymb,amsfonts}
\usepackage{algorithmic}
\usepackage{graphicx}
\usepackage{textcomp}
\usepackage{xcolor}

\def\BibTeX{{\rm B\kern-.05em{\sc i\kern-.025em b}\kern-.08em
    T\kern-.24em\lower.75ex\hbox{E}\kern-.125emX}}
\begin{document}

\title{On the Ability of Deep Learning to Detect Signals with Unknown Parameters\\
}

\author{\IEEEauthorblockN{Tom Anders}
\IEEEauthorblockA{\textit{Wireless@VT} \\
\textit{Virginia Tech}\\
Blacksburg, VA\\
24060}
\and
\IEEEauthorblockN{Hiten Kothari}
\IEEEauthorblockA{\textit{Wireless@VT} \\
\textit{Virginia Tech}\\
Blacksburg, VA\\
24060} 
\and
\IEEEauthorblockN{R. Michael Buehrer}
\IEEEauthorblockA{\textit{Wireless@VT} \\
\textit{Virginia Tech}\\
Blacksburg, VA\\
24060} }

\maketitle

\begin{abstract}
In many signal processing applications, including communications, sonar, radar, and localization,  a fundamental problem is the detection of a signal of interest in background noise, known as signal detection \cite{kay1998fundamentals} \cite{kay2013fundamentals}. A simple version of this problem is the detection of a signal of interest with unknown parameters in Additive White Gaussian Noise (AWGN). When the parameters defining the signal are not known, an optimal detector (in the Neyman-Pearson sense) does not exist. An upper bound on the performance of any detector is the matched filter, which implies perfect sample by sample knowledge of the signal of interest. In recent years Deep Neural Networks (DNNs) have proven to be very effective at hypothesis testing problems such as object detection and image classification. This paper examines the application of DNN-based approaches to the signal detection problem at the raw I/Q level and compares them to statistically based approaches as well as the Matched Filter. These methods aim to maximize the Probability of Detection \(P_d\) while maintaining a constant Probability of False Alarm \(P_{FA}\). Two Machine Learning (ML) algorithms are trained and assessed on this signal detection problem, across three signal of interest models. A model was also trained on a unified dataset and assessed across all signals of interest. 
\end{abstract}


\section{Introduction}
The signal detection problem refers to detecting the presence of a signal that is corrupted by background noise. Knowledge of the presence of a signal of interest is generally a requirement for any further processing. Signal detection performance is a metric of interest in a broad range of RF systems, from communications to radar to radio astronomy. Signal detection techniques are assessed by comparing probability of correctly detecting signals at a specified probability of false alarm \cite{kay1998fundamentals,kay2013fundamentals}. This work aims to characterize the performance of Machine Learning(ML) based classifiers trained for the signal detection task, in order to help assess their viability in future generation RF systems. A detector in the scope of the work presented here is an algorithm that accepts a sequence of samples, and returns a decision as to whether or not a signal of interest is present. Trained ML detectors are compared to a two non-ML baseline detectors, as well as the theoretical upper bound of a Matched Filter (MF).\\

In the case of detection of a signal with precisely known parameters, the well-known upper bound for detection performance is the MF. Oftentimes when the signal of interest is completely unknown, a simple energy detector(ED) is the best approach. On the other hand, if  some characteristics of the signal of interest are known, we can tailor a test to improve upon the ED. The Fisher statistic is a general method that uses thresholding of a calculated quantity to classify into one of the two groups. For the detection of a simple signal, such as a pure tone in noise, we can formulate a Fisher style test in order to gain some advantage over the energy detector.  This approach serves as a useful baseline in the absence of any knowledge of signal parameters. Optimal Fisher statistic derivation becomes exponentially more difficult with increasing signal complexity, so equivalent tests that gain an advantage over the energy detector are not easily formulated. Cyclostationary detectors are not compared to in this work, as it concerns relatively short sequences of interest.\\

Machine learning (ML) methodologies offer the capability of solving mathematically complex optimization problems in efficient ways. Specifically, utilizing ML for the Time Series Classification (TSC) task has been of great research interest in recent years \cite{middlehurst2023bake}. Clearly, the signal detection task falls within time-series classification. Various RF problems have begun to attract deep learning research attention, such as LFM Signal Detection \cite{9023016} and ML-Assisted Signal Detection in Communication Networks \cite{9687463}, among others. 
\cite{9788039} proposes a ML detection framework in an image-processing approach to the spectrogram.  \cite{ahn_machine_2023} describes an ML implemented Signal-to-Noise (SNR) estimator for an OFDM system. \cite{yi_deep_2020} applies ML techniques to the communications signal detection problem, which is related but distinct from the problem examined in this work. \cite{hauser_signal_2017} and \cite{liu_tiny_2022} perform ML based modulation classification utilizing raw I/Q data, a more closely related but still distinct problem. \cite{chavez_dual-layer_2022} utilizes ML architectures for fingerprinting at relatively high SNR, and has been validated over the air. While these are a few examples of similar recent developments, most research on problems that are truly adjacent is reasonably dated \cite{293163}\cite{5685580}\cite{972479}\cite{green1961detection}, especially given advances in Time-Series Analysis \cite{middlehurst2023bake}. 

\section{System Model}
We are concerned with the detection of a signal-of-interest in AWGN, which can be formulated as a binary hypothesis testing problem. We model our system as a set of received raw IQ samples, after downconversion to baseband (represented as the DCV function). 
\begin{equation}
     y[n] = DCV(x[n] + w[n])
\end{equation}
where under the hypothesis ${\cal H}_1$, $x[n]$ is some unknown deterministic signal.
Under the hypothesis ${\cal H}_0$,  $x[n]=0$.  Further,  $w[n]$ are independent samples of Gaussian noise with variance $\sigma^2= \frac{1}{SNR}$ where $SNR$ is the signal-to-noise ratio. We will define a received signal vector 
\begin{equation}
    {\bf y} = \left [ y[0], y[1], \ldots y[N_s-1] \right ]^T
\end{equation} 
where $N_s$ is the number of samples collected.\\

The MF requires perfect knowledge of the signal of interest, and represents a theoretical upper bound for detection. Additionally, we examine two baseline statistical methods: the ED and an FFT thresholding approach \cite{kay1998fundamentals}
\cite{784064}.
Of all detectors that do not require signal knowledge (outside of the SNR), the energy detector is the simplest.  The energy detector serves as a lower bound and simply takes in a received signal vector \(\bf y\) and declares a detection if 
\begin{equation}
    {
\sum_{i=0}^{N_s} |y[i]|^ 2 \lessgtr^{{\cal H}_0}_{{\cal H}_1} \gamma_E}
\end{equation}
where $\gamma_E$ is chosen for the desired $P_{FA}$ and $|\cdot|$ denotes the aboslute value\\

It was previously established that thresholding of the standard periodogram gives the optimum detection performance for a pure tone \cite{784064} with unknown frequency, amplitude, and phase parameters. We can obtain the discrete time approximation of the periodogram by taking the FFT. This process can similarly be represented mathematically as thresholding a calculated quantity against a threshold $\gamma_F$, this time the quantity being based on the FFT of $\mathbf{y}$. Denoting the FFT of $\mathbf{y}$ as $\mathbf{Y}$, a detection is declared if
\begin{equation}
{\frac{\max_{n} |Y[n]|^2}{\sum_{n=0}^{Ns} |Y[n]|^2} \lessgtr^{{\cal H}_0}_{{\cal H}_1} \gamma_F}
\end{equation}
where $\gamma_F$ is again chosen to achieve a constant $P_{FA}$. Equations for calculating this threshold are also presented in \cite{784064}.\\

An upper bound on performance comes from the matched filter, whose decision statistic comes from the correlation of a template of known expected samples with the received samples.
\begin{equation}
    r = \sum_{n=0}^{N_s -1} y[n] * h^*[n]
\end{equation}
Where $h^*[n]$ denotes the elementwise complex conjugate of template $h[n]$ and $\gamma_M$ is chosen for a desired $P_{FA}$.
\begin{equation}
\frac{r}{N_s} \lessgtr^{{\cal H}_0}_{{\cal H}_1} \gamma_M
\end{equation}

Three different signals of increasing complexity were used for performance assessment. These were a pure sine tone, QPSK, and OFDM. Each of these signals were simulated as a real signal with a center frequency of \(f_{True} = 75\)KHz and then downconverted to baseband using an estimated center frequency uniformly distributed between \(74\)-\(76\)KHz, resulting in a leftover frequency offset of \(\pm 1\)KHz. Datasets were generated and performance was assessed on each of these signals of interest. \\

Datasets were generated for each signal type, made up of 70,000 sample sequences apiece, 35,000 of noise only, and 35,000 containing a signal of interest in noise, with 1,000 sample sequences for each SNR bin from -30 to 5dB. Validation sets were identically structured. Other relevant simulation parameters are shown in Table I. For example each sequence was 500 samples long.
\begin{table}[h!]
\caption{Simulation Parameters}
\begin{center}
\begin{tabular}{|c|c|c|} 
\hline
{Symbol}&{Variable Name}&{Value Used} \\
\hline
\hline
\multicolumn{3}{|c|}{\textbf{General}}\\
\hline
{\(f_s\)}&{Sampling Frequency}&{\(2.048\)MHz }\\
\hline
{\(f_c\)}&{Estimated Frequency}&{74-76KHz} \\
\hline
{\(f_{True}\)}&{True Frequency}&{75KHz} \\
\hline
{\(\Delta f\)}&{Frequency Offset}&{-1 to 1 KHz}\\
\hline
{\(\phi\)}&{Phase Offset}&{\(-\pi\) to \(\pi\)}\\
\hline
{\(N_s\)}&{Samples/Sequence}&{500 }\\
\hline
{\(SNR\)}&{Signal to Noise Ratio}&{\(-30\) to \(5\) dB}\\
\hline
{\(Samples\)}&{Sample Sequences/Bin}&{\(1000\) }\\
\hline
{\(LPF\)}&{Filter Used}&{Butterworth }\\
\hline
{\(\omega_c\)}&{Bandwidth}&{\(40000\)Hz }\\
\hline
\hline
\multicolumn{3}{|c|}{\textbf{QPSK}}\\
\hline
{\(f_{sym}\)}&{Symbol Rate}&{25KHz }\\
\hline
{\(\alpha\)}&{Roll-Off Factor}&{0.4 }\\
\hline
\hline
\multicolumn{3}{|c|}{\textbf{OFDM}}\\
\hline
{}&{Inner Modulation}&{QPSK}\\
\hline
{\(f_{sym}\)}&{Symbol Rate}&{32000}\\
\hline
{SCS}&{Sub-carrier Spacing}&{2000}\\
\hline
{\(N_c\)}&{Number of Carriers}&{16}\\
\hline

\end{tabular}
\label{tab1}
\end{center}
\end{table}

Sequences were generated as purely real, at which point noise was based on SNR bin before downconversion. The downconversion process was defined by:
\[DCV(\boldsymbol{x}) = LPF(\boldsymbol{x} * cos(\boldsymbol{n})) + 1j * LPF(\boldsymbol{x} * sin(\boldsymbol{n}))\]
Where \[\boldsymbol{n} = [1,...,N_s] * 2 \pi f_{c} / f_{s}  \] and LPF represents low pass filtering. Note that $f_c$ is here drawn from a uniform distribution between 74KHz-76KHz, representing an up to 1KHz error in estimated center frequency. For data used for training and evaluating ML methods, downconversion was followed by a normalization defined by: \[\mathbf{x}_{norm} = \frac{\mathbf{x}}{\max_{i} |x[i]|}\]
This normalization was practical for ML methods as it constrained all samples to a maximum magnitude of 1, preventing learning based on magnitude bounds. \\

Models from the TSAI(Time-Series AI) python repository were used for assessment\cite{tsai}. The repository offers streamlined access to GPU-ported versions of many different state of the art TSC implementations. Different TSC models performed similarly on the signal detection task, so high performing models with reasonable training and inference times, MiniRocketPlus \cite{dempster_minirocket_2021}, InceptionTimePlus \cite{fawaz_inceptiontime_2020} were selected for assessment. These models learn directly from raw in-phase and quadrature (I/Q) data without the need for explicit interference templates.

\subsection{InceptionTimePlus}
InceptionTimePlus is a deep convolutional model built on top of InceptionTime introduced by Fawaz et al. \cite{inception_time}. It is built using stacked Inception modules, each consisting of parallel 1D convolutional filters with different kernel sizes to capture both short and long term dependencies in the input time series. To reduce computational overhead and mitigate overfitting, a 1x1 bottleneck convolution is applied before the larger kernels. It also employs batch normalization and dropout for better performance in terms of stability and overfitting.

\subsection{MiniRocketPlus}
MiniRocketPlus is a time series classification model that builds upon the highly efficient and scalable MiniRocket framework introduced by Dempster et al. \cite{Dempster_2021}. Unlike deep neural network-based approaches, MiniRocketPlus leverages convolutional feature extraction using deterministic kernels and a simple linear classifier, offering a good balance between accuracy and computational efficiency. MiniRocket uses Proportion of Positive Values (PPV) pooling to compute a single feature for each of the resulting feature maps (i.e., the proportion of positive values). The transformed features are used to train a linear classifier, which is extremely fast to train and deploy.

MiniRocketPlus is a generally faster algorithm, but these both are high performers on various TSC tasks.  

\section{Simulation and Results}

Three versions of each classifier were trained, with the following training datasets:
\begin{enumerate}
    \item Sinusoid
    \item QPSK
    \item OFDM
\end{enumerate}

Each of the trained classifiers was assessed on an independent validation dataset. The MF upper bound and Fisher Statistic are also shown for each dataset. It can be seen that while the Fisher Statistic performs better than the Energy Detector for the Sine Wave detection problem, its advantage falls off for QPSK and OFDM, in which energy is much more spread out. Probability of Detection \(P_d\) was assessed on fully separated validation datasets. Thresholds were set for ML classifiers in order to achieve the desired \(P_{FA}\) of 0.01 within \(\pm 0.001\). Figures 1-3 show \(P_d\) vs SNR (dB) for the three training datasets. It can be seen that the Inception classifier on the Sine dataset offered close to 3dB advantage over FFT Thresholding. Similarly, for QPSK, The superior classifier was again Inception, but this time it only barely beat out the ED method. However, neither classifier was able to outperform the ED on the OFDM dataset. Overall, as signal complexity increased, performance of the ML classifiers was diminished. \\

The Inception model was also trained on a unified dataset of all three signals-of-interest, and tested on their ability to recognize the presence of any of the three. Figure 4 shows performance for this case. Energy Detector performance was the same for each signal of interest, so it is only plotted once. Training on the unified dataset resulted in very small (\(<\)1dB) performance losses vs training several models on individual datasets.

\begin{figure}[h!]
    \centering
    \includegraphics[scale=0.22]{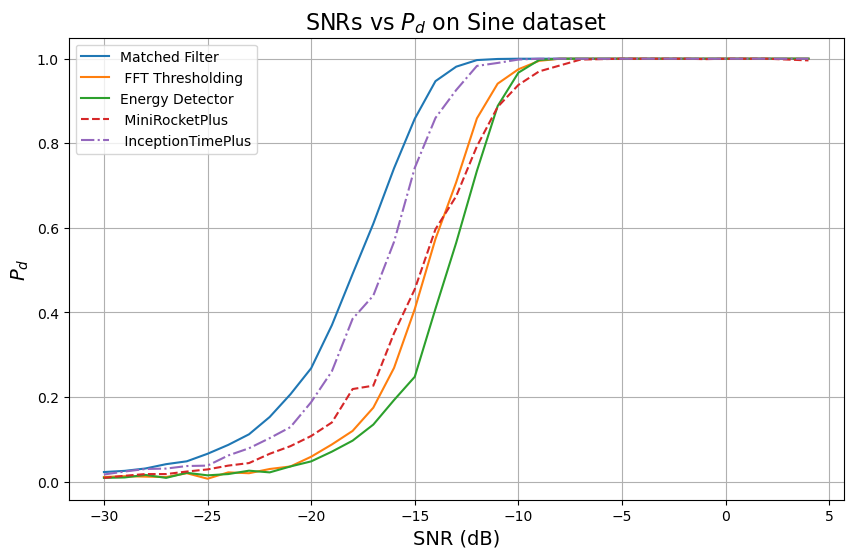}
    \caption{Sine Wave Performance}
    \label{fig:enter-label}
\end{figure}

\begin{figure}[h!]
    \centering
    \includegraphics[scale=0.22]{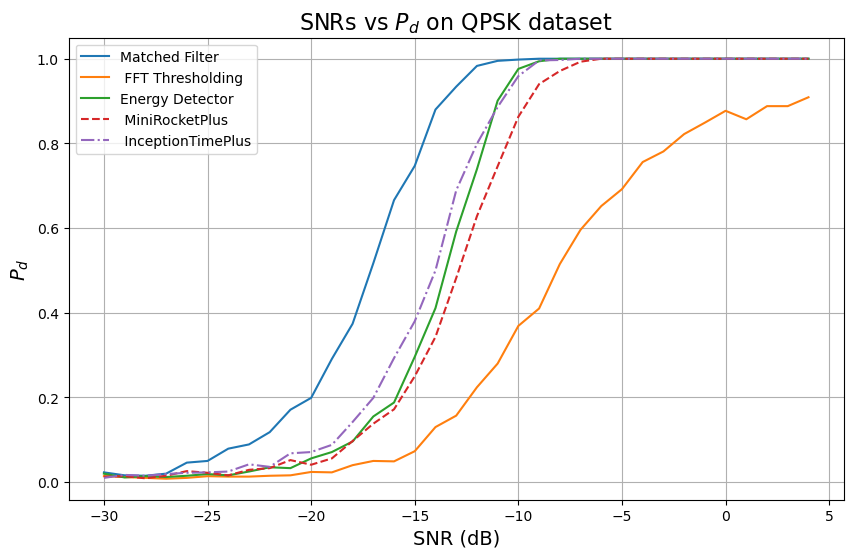}
    \caption{QPSK Performance}
    \label{fig:enter-label}
\end{figure}

\begin{figure}[h!]
    \centering
    \includegraphics[scale=0.22]{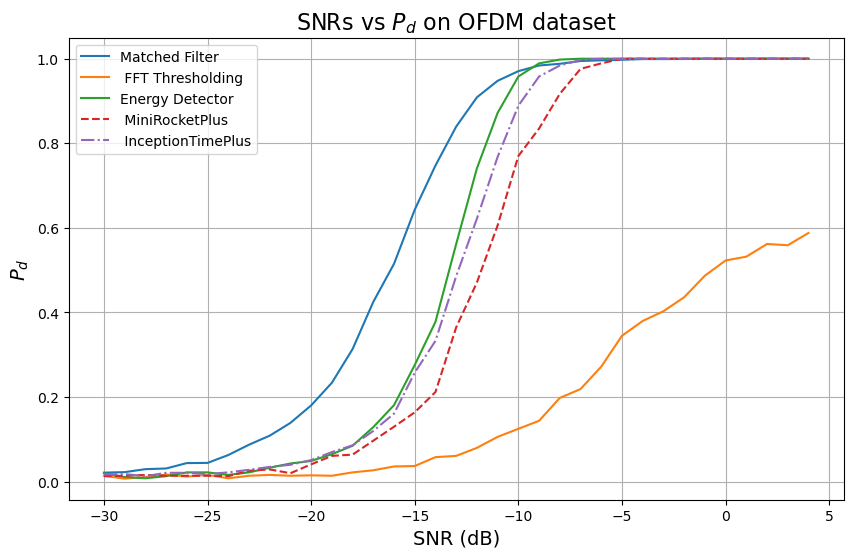}
    \caption{OFDM Performance}
    \label{fig:enter-label}
\end{figure}
\begin{figure}[h!]
    \centering
    \includegraphics[scale=0.22]{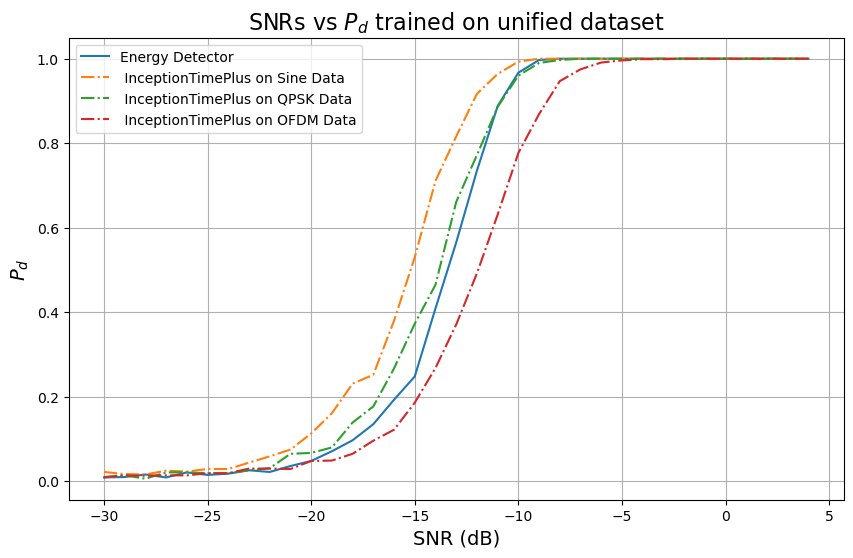}
    \caption{Universal Classifier Performance}
    \label{fig:enter-label}
\end{figure}

\section{Conclusions}
Modern ML methods provide unique adaptability to mathematically and computationally complex problems. Provided enough examples, ML  algorithms can numerically find well-performing solutions to potentially extremely complex mathematical optimization problems. This can be useful in situations where we can simulate expected behavior,  but can't necessarily solve for the black box. For example, when we have imperfect knowledge of signals of interest (such as true carrier frequency), datasets can be created that accurately reflect the impact of unknowns and pass this knowledge onto models that are trained on them. Notably, no extensive optimization was performed on model parameters from the TSAI toolkit, so there may be fine tuning that can be done to squeeze out performance.\\

In some RF applications, it may be that real-time suitable computational complexity is prioritized over best possible detection performance, but each situation has different requirements and a different optimal solution. It should be noted that while ML solutions will be of higher complexity than statistically derived methods, hardware suitable for ML applications is becoming more accessible, faster, and more power efficient.\\

Clearly, improving ML performance on more complex signals has potential as a future thrust area. This might be accomplished through more powerful classification models or well-designed pre-processing in order to make it easier for models to recognize features. One parameter of interest not examined here is number of samples per sequence. In general, as this increases, all detectors perform better, but ML methods take longer for training and inference. Preliminary efforts have also taken place towards applying this work to signals collected over-the-air. Initial results were encouraging, despite the numerous limitations of quantifying over the air performance. An important consideration in future work that aims to extend ML detection to over-the-air data should be the use of noise models realistic to the application in training data. Another natural extension to the signal detection problem is parameter estimation, which could set the stage for performing more complex processing tasks such as interference mitigation using ML methods. 

\section*{Acknowledgement}
This paper is based upon work supported in part by the National Science Foundation under Grant ECCS-2029948.

\printbibliography
\end{document}